\documentclass[11pt,twoside]{article}
\usepackage{asp2010}

\resetcounters

\begin{document}

\title{On the Physics of Kinetic-Alfv\'en Turbulence}
\author{Stanislav Boldyrev$^1$ and Jean Carlos Perez$^{1,2}$
\affil{$^{1}$Department of Physics, University of Wisconsin, 1150 University Ave, Madison, WI 53706, USA}
\affil{$^2$Space Science Center, University of New Hampshire, Durham, NH 03824, USA}
}

\begin{abstract}
Observations reveal nearly power-law spectra of magnetic and density plasma fluctuations at the subproton scales in the solar wind, which  indicates the presence of a turbulent cascade. We discuss the three-field and two-field models for micro-scale plasma fluctuations, and then present the results of numerical simulations of a two-field model of kinetic-Alfv\'en turbulence, which models plasma motion at  sub-proton scales. 

\end{abstract}

\section{Introduction}
Magnetic turbulence is ubiquitous in astrophysical 
systems and it is present in laboratory devices.  Turbulence may be naturally generated due to various instabilities (such as supernovae explosions and galactic shear in the interstellar medium, tearing modes and shear flows in laboratory devices. Nonlinear energy cascade transfers the energy to smaller and smaller scales, thus distributing turbulent energy over a broad range of scales. At scales much larger than plasma microscales (ion cyclotron radius, skin depth, etc), fundamental properties of plasma turbulence can be understood in the framework of magnetohydrodynamics (MHD). Analytic and numerical studies of MHD turbulence allowed one to explain qualitatively and, in some cases, quantitatively in situ observations of plasma turbulence in the solar wind \cite[e.g.,][]{boldyrev_etal2011,wang2011,zhdankin_etal2012}. 

When the energy reaches the scales comparable to the ion Larmor radius, the character of the turbulence changes.  In laboratory flusion plasmas such micro-turbulence is responsible for transport phenomena. Since the large-scale guide magnetic field is typically strong and it cannot be easily perturbed in fusion devices, the studies have been mostly devoted to electrostatic fluctuations. Recently, there appeared reliable in situ measurements of sub-proton plasma turbulence in the solar wind, where magnetic fluctuations are essential \cite[e.g.,][]{alexandrova_etal2009,alexandrova_etal2011,sahraoui_etal2009,chen_etal2010,chen_etal2012a,salem_etal2012}. Such small-scale turbulence is thought to be responsible for energy dissipation and plasma heating in the solar wind. 

A major possibility is that significant role in subproton turbulence is played by kinetic-Alfv\'en modes. Indeed, one can argue that the cascade of strong MHD turbulence (that is, turbulence of shear Alfv\'en modes whose linearized dispersion has the form $\omega\propto k_zv_A$) is expected to transform into the cascade of kinetic-Alfv\'en turbulence (whose linearized dispersion relation is $\omega \propto k_zk_\perp$) at subproton scales \cite[e.g.,][]{schekochihin_etal2009}.  The argument goes as follows. Typical frequency of Alfv\'enic fluctuations is $\omega \approx k_\| v_A$. The anisotropy of the energy distribution implies $k_\|\ll k_\perp$ at small scales. At the proton gyroscale, $k_\perp \sim 1/\rho_i$, one therefore estimates $\omega \ll k_\perp v_A \sim k_\perp v_{Ti}\sim \Omega_i$, where $v_{Ti}$ is the thermal ion velocity, $\Omega_i$ is ion gyrofrequency, and we assumed plasma beta of order one, that is $v_A\sim v_{Ti}$. Therefore, at proton gyroscales, turbulence frequency is expected to be smaller than the ion gyrofrequency so that anisotropic kinetic Alfv\'en modes may be effectively generated. At present, subproton kinetic-Alfven turbulence is understood to a significantly lesser extent compared to its Alfv\'enic MHD counterpart. 

In this contribution we discuss a fluid model for micro-scale plasma fluctuations, and then present the results of numerical simulations of of kinetic-Alfv\'en turbulence, which help to explain recent measurements of magnetic and density fluctuations at the subproton scales in the solar wind.

\section{Kinetic-Alfv\'en Equations}
The  equations governing kinetic-Alfv\'en turbulence have been derived and studied in many works  \cite[e.g.,][]{hazeltine1983,scott_hd1985,camargo_etal1996,terry_mf2001,schekochihin_etal2009,smith_t2011,boldyrev_p12}. The basic assumptions are that a uniform background magnetic field is strong compared to magnetic fluctuations, $B_0\gg b$, and the turbulence is strongly anisotropic, $k_z\ll k_\perp$, where $k_z$ and $k_\perp$ are typical wavenumbers of turbulent fluctuations in the field-parallel and field-perpendicular directions.  To illustrate the essential physics and to set the notation, we start with the simplest case of small plasma beta (the ratio of thermal plasma energy to the magnetic energy, $\beta=8\pi n_0T/B_0^2$). Then we demonstrate how the derived system of equations can be extended for for the case of $\beta \sim 1$, which is relevant for the solar wind stidues.

We will assume the fluid description for the electrons. 
The electrons are advected across the magnetic field by the ``E cross B'' drift, ${\bf v}_{e\perp}= c{\bf E}\times {\bf B}_0/B_0^2$, while their field-parallel motion is related to the current $J_\|= - e n_e v_{e \|}$, and the ion parallel motion can be neglected. For small $\beta$ the fluctuations of the magnetic field strength can be neglected, that is, the magnetic fied is represented as ${\bf B}=B_0{\hat z}+{\bf b}_\perp$. The field-perpendicular component is expressed through the flux function ${\bf b}_\perp = {\hat z} \times \nabla \psi$, so that $J_\|\approx J_z =(c/4\pi)\nabla_\perp \times {\bf b}_\perp= (c/4\pi)\nabla_\perp^2 \psi$.  The flux function is the (minus)   field-parallel component of the vector potential, $\psi= - A_z$. 

The field-parallel force balance in the electron momentum equation gives $-  \nabla_\| (p_e) -n_0 e{\bf E}_\|= 0$, where the electric field is ${\bf E}=-\nabla \phi -(1/c)\partial_t {\bf A}$. Supplementing this equation with the electron continuity equation, one obtains the system for the fluctuating parts of magnetic and density fields:\footnote{When derivatives are taken, we must distinguish the gradients along the guide field ${\bf B}_0$, $\nabla_z$, from the gradients along the local field ${\bf B}={\bf B}_0+{\bf b}$, $\nabla_\|$.}
\begin{eqnarray}
& \frac{1}{c}\frac{\partial}{\partial t} {\psi}-\nabla_\| {\phi} +\frac{1}{n_0 e}\nabla_\|{p_e}= 0, 
\label{psi}\\
& \frac{\partial}{\partial t} { n_e}  - \frac{c}{B_0}\nabla {\phi}\times {\hat z}\cdot \nabla {n_e}- \frac{1}{e}\nabla_\| {J}_\|=0. 
\label{n}
\end{eqnarray}
The electron equation can be further simplified since the electron thermal speed exceeds the Alfv\'en speed, and an isothermal fluid description is possible, $T_e={\rm const}$. This condition applies for a collisionless plasma, and it also requires not too small plasma beta, $\beta > m_e/m_i$. When collisions cannot be neglected, the electron fluid is isothermal if the electron diffusion time in the field-parallel direction $\tau_{diff}\sim 1/(k_\|^2 v_{Te}^2\tau_{coll})$ is less than the inverse frequencies of corresponding plasma fluctuations. 

The smallness of the plasma beta is essential for neglecting the fluctuations of the magnetic field strength. In the case of $\beta \sim 1$, the fluctuations of the magnetic-field strength cannot be neglected, and the magnetic field is represented as ${\bf B}=(B_0+b_{z}){\hat z}+{\bf b}_\perp$. The fact that the z-component of magnetic fluctuations should be retained follows from the field-perpendicular force balances in the ion and electron momentum equations, which can be combined to give: $-\nabla_\perp p_e -\nabla_\perp p_i +(1/c){\bf J}\times {\bf B}=0$. We therefore derive
\begin{eqnarray}
-\frac{1}{4\pi}B_0\nabla_\perp b_z-\nabla_\perp p_e -\nabla_\perp p_i=0,
\label{force_balance}
\end{eqnarray}
which gives an estimate  ${b_z}/{B_0} \sim \beta ({n_e}/{n_0}).$ For $\beta \ll 1$, the fluctuations of $b_z$ can be neglected, while for $\beta \sim 1$ they should be retained. 

It is not difficult to modify equations (\ref{n}) and (\ref{psi}), taking into account $b_z$. The modification comes in two ways. First, the ``E cross B" velocity should be modified by taking into account $b_z$,
\begin{eqnarray} 
{\bf v}_{e\perp}= c{\bf E}\times ({\bf B}_0+b_z{\hat z})/B^2,
\label{ve}
\end{eqnarray}
where $B^2\approx B_0^2+2B_0b_z$. Second, in the electron continuity equation (\ref{n}) one has to take into account the diamagnetic drift velocity,
\begin{eqnarray} 
{\bf v}_{e*}= \frac{c}{neB^2}{\nabla p_e}\times ({\bf B}_0+b_z{\hat z}).
\label{v*}
\end{eqnarray}
This step requires an explanation. When the magnetic field strength does not change, that is, $b_z=0$, it can be checked that the diamagnetic drift does not advect the electron density. Physically, this happens because guide centers of particles do not move when the diamagnetic current is present. That is why the diamagnetic drift does {\em not} enter Eq.~(\ref{n}) even in the case of a general equation of state. However, if the magnetic field strength changes, the magnetic curvature effects do affect the density advection, and terms with derivatives of $b_z$ do not cancel out.

Straighforward substitution of the modified drift velocities (\ref{ve}) and (\ref{v*}) into the electron  continuity equation then gives the modified equation (\ref{n}):
\begin{eqnarray}
\frac{\partial}{\partial t} \left[\frac{ n_e}{n_0}-\frac{b_z}{B_0} \right]  - \frac{c}{B_0}\nabla {\phi}\times {\hat z}\cdot \nabla \left[ \frac{n_e}{n_0}- \frac{b_z}{B_0}\right] - \frac{c}{eB_0}\nabla \left(\frac{p_e}{n_0}\right)\times {\hat z}\cdot \nabla \left(\frac{b_z}{B_0}\right)- \frac{1}{en_0}\nabla_\| {J}_\|=0, 
\label{nmodified}
\end{eqnarray}
which is derived for an arbitrary $p_e$ but will be simplified using the isothermal equation of state \cite[cf. ][Eq.~(C7)]{schekochihin_etal2009}. 
In the limit of small plasma beta we have $b_z \to 0$, and Eq.~(\ref{nmodified}) turns into Eq.~(\ref{n}). 
We should note that the field-parallel gradient in these equations is the gradient along the total magnetic field, that is,
\begin{eqnarray}
\nabla_\|=\nabla_z + \frac{1}{B_0}{\hat z} \times \nabla {\psi} \cdot \nabla \,.
\label{nabla}  
\end{eqnarray}
In our discussion of strong kinetic Alfv\'en turbulence, we will assume that the fluctuations are anisotropic with respect to the magnetic field in such a way that the so-called critical balance between the linear and nonlinear terms is satisfied, $B_0\nabla_z \sim {\hat z} \times \nabla { \psi} \cdot \nabla$; this condition is analogous to $k_z B_0 \sim k_\perp  b$  \cite[e.g.,][]{goldreich_s95,cho_l2004,howes_etal2011b,tenbarge_h2012}. When this condition is satisfied, equations (\ref{psi}, \ref{n}) or (\ref{psi}, \ref{nmodified}) are essentially nonlinear and three-dimensional.

The system (\ref{psi}, \ref{n}) involving the three fields, $n_e$,  $\psi$, and $\phi$ or the system (\ref{psi}, \ref{nmodified}) involving the four fields $n_e$, $b_z$, $\psi$, and $\phi$ are incomplete, as they have more independent fields than equations. The uniqueness is restored when the systems are  supplemented by the equations for the ions. The situation here depends on the scales considered. Above the ion-cyclotron scale $\rho_i=v_{Ti}/\Omega_i$, a fluid description can be justified for the ions if the ions are cold, which is essentially the limit of low beta. (This case is applicable for most laboratory plasmas, where the corresponding equations have been originally derived.) In this case the ions move across the magnetic field due to ``E cross B'' drift and the polarization drift, and one can write the charge concervation law, $ \partial \rho /{\partial t} +\nabla_\perp {\bf J}_\perp +\nabla_\| J_\|=0$, 
where the parallel current is given by the electrons, while the perpendicular current is due to the polarization drift of the ions (the ``E cross B'' drifts are the same for ions and electrons, they do not lead to charge separation and do not contribute to the current). The resulting equation is \cite[e.g.,][]{terry_mf2001}:
\begin{eqnarray}
\frac{n_im_ic^2}{B_0^2}\left[\frac{\partial}{\partial t}\nabla^2\phi -\frac{c}{B_0}\nabla \phi \times {\hat z}\cdot \nabla \nabla^2\phi \right]=\nabla_{\|} J_{\|}. 
\label{phi}
\end{eqnarray}
Equations (\ref{psi}), (\ref{n}), and (\ref{phi}) provide the closed {\em three-field} system for evolution of electric, magnetic and density fields in the case of low plasma beta. 

If plasma beta is not small, the ions require kinetic description, and a simple fluid model is not well justified. We however will be interested in the sub-proton, dispersive kinetic-Alfv\'en waves, that is, we consider scales smaller than the ion gyroscale $k_\perp \rho_i \gg 1$. It is also convenient to introduce the ion-acoustic scale, $\rho_s=v_s/\Omega_{i}$ with $v_s=(T_e/m_i)^{1/2}$ the ion acoustic speed. At such scales, the ions are (spatially) not magnetized. Moreover, we will be interested in frequencies smaller than $kv_{Ti}$, which implies the ``Boltzmannian'' response for the ion density fluctuations, $n_i=- e \phi n_0/T_i$. Note that we do {\em not} require the frequencies to be smaller than the ion gyrofrequency, as it is implied, e.g., in gyrokinetic treatments.   
The quasi-neutrality condition $n_i = n_e$ then relates the electric potential to the electron density, $ \phi =-(T_i /n_0 e) n_e$. Similarly, in the three-field system, $b_z$ field can be removed from (\ref{nmodified}) according to Eq.~(\ref{force_balance}): $b_z=-4\pi (T_i+T_e)n_e/B_0$.

\section{Kinetic-Alfv\'en Turbulence}
Let us introduce the normalized electron density and the magnetic flux function,
\begin{eqnarray}
{\tilde n}={(1+T_i/T_e)^{1/2}(v_s/v_A)}{\left[1+(v_s/v_{A})^2(1+T_i/T_e)\right]^{1/2}} \frac{n_e}{n_0}, \quad {\tilde \psi}=\frac{v_s e}{c T_e} \psi,
\end{eqnarray}
and normalize the time and the length according to
\begin{eqnarray}
{\tilde t}=\frac{\left(1+{T_i}/{T_e}\right)^{1/2}}{\left({\rho_s}/{v_A}\right)\left[ 1+({v_s}/{v_A})^2\left(1+{T_i}/{T_e} \right)  \right]^{1/2}}t, \quad {\tilde {\bf x}}={\bf x}/\rho_s .
\end{eqnarray}
We will use only the normalized variables (unless stated otherwise) and omit the over-tilde sign. The system (\ref{psi}), (\ref{nmodified}) then takes the form:
\begin{eqnarray}
& \partial_t {\psi} + \nabla_\|{n}= 0, 
\label{psimod}\\
& \partial_t {n} -\nabla_\| \nabla_\perp^2 \psi=0,  
\label{nmod} 
\end{eqnarray}
where $\nabla_\|=\nabla_z+ {\hat z} \times \nabla {\psi} \cdot \nabla_\perp$. The presented ideal system conserves the total energy $E$ and the cross-correlation $H$,
\begin{eqnarray}
& E=\int \left( |\nabla_\perp \psi|^2 + n^2 \right)d^3 x, 
\label{energy}\\
& H=\int \psi n d^3 x .
\end{eqnarray}
The system (\ref{psimod},\ref{nmod}) possesses linear waves, $n_k\propto \psi_k \propto \exp(-i\omega t +i{\bf k}{\bf x})$. The linearization is done by neglecting the second term in the right-hand side of Eq.~(\ref{nabla}), which gives the dispersion relation for the kinetic-Alfv\'en waves: 
\begin{eqnarray}
\omega= k_z k_\perp .
\label{dispersion}
\end{eqnarray}
The linear modes are characterized by the equipartition between the density and magnetic fluctuations, $n_k=\pm k_\perp \psi_k$.

\begin{figure}[tbp!]
\hskip15mm \includegraphics[width=0.7\columnwidth]{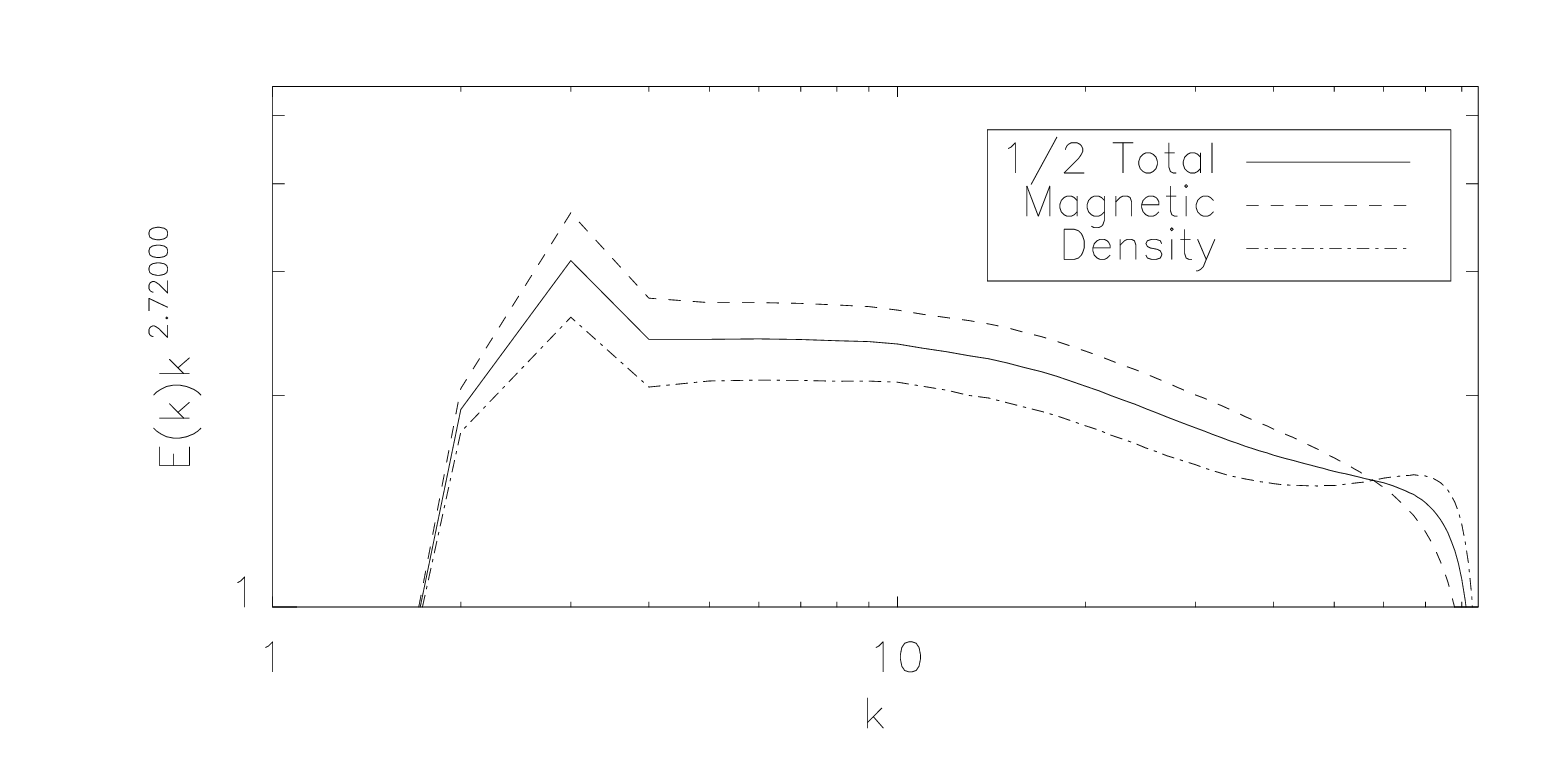}
\caption{\label{fig:spectrum} Energy spectrum of strong kinetic-Alfv\'en turbulence at sub-proton scales, obtained in two-field numerical simulations with spatial resolution $256^3$.}
\end{figure}

For numerical simulations we supplement the equations with large-scale random forces that supply the energy to the system:
\begin{eqnarray}
& \partial_t {\psi} + \nabla_\|{n}= \eta \nabla_\perp^2 \psi + f_{\psi}, 
\label{eq1}\\
& \partial_t {n} - \nabla_\| \nabla_\perp^2 \psi = \nu \nabla_\perp^2 n + f_n. 
\label{eq2}
\end{eqnarray}
The small dissipation terms serve to remove the energy at small scales (and they are mostly needed to stabilize the code). In a turbulent state, the energy cascades toward small scales while the cross-correlation cascades toward large scales. The numerically obtained energy spectrum is shown in Fig.~\ref{fig:spectrum}. It is steeper than the spectrum $-7/3$ predicted by phenomenological theories based on dimensional arguments \cite[e.g.,][]{biskamp_etal1999,cho_l2009}. 
It is interesting that a spectrum steeper than $-7/3$ was also inferred from observations of subproton magnetic and density fluctuations in the solar wind \cite[e.g.,][]{chen_etal2010,alexandrova_etal2011,chen_etal2012a}.

Various explanations have been proposed for the steeper than $-7/3$ spectrum of subproton turbulence observed in the solar wind. They include steepening of the spectrum by Landau damping, weakening of turbulence, wave-particle interactions, etc. \cite[e.g.,][]{rudakov_etal2011,howes_etal2011b}. In our model wave-particle interactions are absent, however, the steeper spectrum persists. A possible  explanation proposed in \cite{boldyrev_p12} invoked intermittency corrections that result from two-dimensional structures formed by density and magnetic fluctuations. It was proposed that the spectrum should be close to $-8/3$, the value consustent with observations and numerical simulations. This points to an interesting possibility that the observed scaling is not an artifact of non-universal or dissipative effects, rather, it is an inherent property of the nonlinear turbulent dynamics. The spectrum may therefore be universal, analogous to the Kolmogorov spectrum of fluid turbulence. A definitive numerical study that requires higher numerical resolution will be conducted elsewhere.

\acknowledgments 
This work was supported   
by the US DOE Awards DE-FG02-07ER54932, DE-SC0003888, and DE-SC0001794, the NSF Grant PHY-0903872, the NSF/DOE Grant AGS-1003451, and by the NSF Center for Magnetic Self-organization in Laboratory and Astrophysical Plasmas at the University of Wisconsin-Madison.


\end{document}